\documentclass[%
 reprint,
 amsmath,amssymb,
]{revtex4}
\usepackage[section]{placeins}
\let\Oldsection\section
\renewcommand{\section}{\FloatBarrier\Oldsection}

\let\Oldsubsection\subsection
\renewcommand{\subsection}{\FloatBarrier\Oldsubsection}

\let\Oldsubsubsection\subsubsection
\renewcommand{\subsubsection}{\FloatBarrier\Oldsubsubsection}
\usepackage{graphicx}
\usepackage{dcolumn}
\usepackage{bm}

\begin{document}

\preprint{APS/123-QED}
\title{A degenerate three-level laser coupled to a squeezed vacuum reservoir}

\author{Dawit Hiluf}
\email{dawit.hailu@huji.ac.il}
\affiliation{Physics Department, Mekelle University, P.O.Box 231, Mekelle, Ethiopia.}
\author{Fesseha Kassahun}
\email{fessehakassahun@gmail.com }
\affiliation{Department of physics, Addis Ababa University, P.O.Box 1176, Addis Ababa, Ethiopia}

\date{\today}
\begin{abstract}
Employing the master equation for a three-level laser driven by coherent light and coupled to a squeezed vacuum reservoir, we obtain stochastic differential equations associated with the normal ordering. Using the solutions of the stochastic differential equations, we calculate the quadrature variance, the squeezing spectrum, the mean photon number, and the variance of the photon number. It turns out that the degree of squeezing increases with the linear gain coefficient or the squeeze parameter. It is also found that the driving coherent light decreases the mean photon number.

\begin{description}
\item[Usage]
Secondary publications and information retrieval purposes.
\item[PACS numbers]
May be entered using the \verb+\pacs{#1}+ command.
\item[Structure]
You may use the \texttt{description} environment to structure your abstract;
use the optional argument of the \verb+\item+ command to give the category of each item. 
\end{description}
\end{abstract}

\pacs{Valid PACS appear here}
\maketitle


\section{Introduction}
  In quantum optics, the annihilation and creation
operator describing a single-mode radiation can be decomposed into
two component operators, referred to as quadrature operators. For
a single-mode radiation in any state, the product of the
fluctuations in the two quadratures satisfies the uncertainity
relation ~\cite{1}. In a squeezed state the quantum noise in one
quadrature is below the vacuum level at the expense of enhanced
fluctuations in the conjugate quadrature, with the product of the
variance in the two quadratures satisfying the uncertainty
relation. Squeezing like photon antibunching or sub-Poissonian
photon statistics is a nonclassical feature of light ~\cite{2}.
Parametric oscillation and second harmonic generation are typical
processes leading to the production of squeezed light modes
~\cite{2,3,4}. It is also worth mentioning that squeezed light has
potential applications in the detection of weak signals, noiseless
communication, and precision measurement ~\cite{2,4,5,6}.

 The squeezing and statistical properties of the light
produced by a degenerate three-level laser coupled to a vacuum
reservoir have been investigated by several authors when either
the atoms are initially prepared in a coherent superposition of
the top and bottom levels ~\cite{7,8,9}, or when these levels are
coupled by a strong coherent light ~\cite{10,11,12}. Moreover, the
squeezing and statistical properties of the light produced by a
degenerate three-level laser coupled to a squeezed vacuum
reservoir and in which the atoms injected into the cavity are
initially prepared in a coherent superposition of the top and the
bottom levels have been investigated recently~\cite{14}. It is
found that a three-level laser generates under certain condition
squeezed light ~\cite{2,7,8,9,10,11,12,13,14,15,16}.

The main objective of this paper is to analyze the squeezing and
statistical properties of the light produced by a three-level
laser coupled to a squeezed vacuum reservoir via a single-port
mirror  and in which the top and bottom levels of the three-level
atoms injected into the cavity are coupled by a strong coherent
light. We carry out the analysis  of this quantum optical system
using the pertinent stochastic differential equations for the
cavity mode variables, associated with the normal ordering. The
solution of the resulting equations are then used to calculate the
quadrature variance, the squeezing spectrum, the mean photon
number, and the variance of the photon number.

\section{Stochastic Differential Equationsl}
\label{sec:thesystem}
 A three-level laser consists of a cavity
into which three-level atoms in a cascade configuration are
injected at a constant rate~$r_a$~and removed from the cavity
after a certain time~$\tau.$~We represent the top, middle, and
bottom levels by
~$|a\rangle,$~~$|b\rangle,$~and~$|c\rangle,$~respectively. In
addition, we assume the cavity mode to be at resonance with the
two transitions
~$|a\rangle\rightarrow|b\rangle$~and~$|b\rangle\rightarrow|c\rangle,$~with
direct transition between levels~$|a\rangle$~and~$|c\rangle$~to be
dipole forbidden ~\cite{2}. We consider the case in which the
atoms are initially prepared in a superposition of the top and
bottom levels and in which these levels are coupled by a strong
coherent light. The Hamiltonian describing the coupling between
levels~$|a\rangle$~and~$|c\rangle$~by the  coherent light can be
expressed at resonance as
$$ \hat{H'}=ig'(\hat{b}^\dag|c\rangle\langle
a|- \hat{b}|a\rangle\langle c|),\eqno(2.1)$$where ~$g'$~is the
coupling constant and~$\hat{b}$~is the annihilation operator for
the coherent light. Assuming the coherent light to be strong, we
replace~$\hat{b}$~by~$\beta,$~which is taken to be real, positive,
and constant. In view of this, we can write the Hamiltonian as
$$ \hat{H'}={i\Omega\over 2}( |c\rangle\langle
a|-|a\rangle\langle
c|),\eqno(2.2a)$$where~$$\Omega=2g'\beta\eqno(2.2b)$$~is called
the Rabi frequency. On the other hand, the Hamiltonian describing
the interaction between a three-level atom and the cavity mode can
be written at resonance as
$$ \hat{H}''=ig[\hat{a}^\dag(|b\rangle\langle a|+|c\rangle\langle
b|) -\hat{a}(|a\rangle\langle b|+|b\rangle\langle
c|)],\eqno(2.3)$$where~$\hat{a}$~is the annihilation operator for
the cavity mode. On account of Eqs. (2.2a) and (2.3), the
Hamiltonian describing the interaction of a three-level atom with
the coherent light and the cavity mode is
$$ \hat{H}={i\Omega\over 2}(|c\rangle\langle
a| -|a\rangle\langle c|)+ig[\hat{a}^\dag(|b\rangle\langle
a|+|c\rangle\langle b|)-\hat{a}(|a\rangle\langle
b|+|b\rangle\langle c|) ].\eqno(2.4)$$
\begin{center}
\begin{figure}[htp]
\centerline{\includegraphics [width=2.5 in]{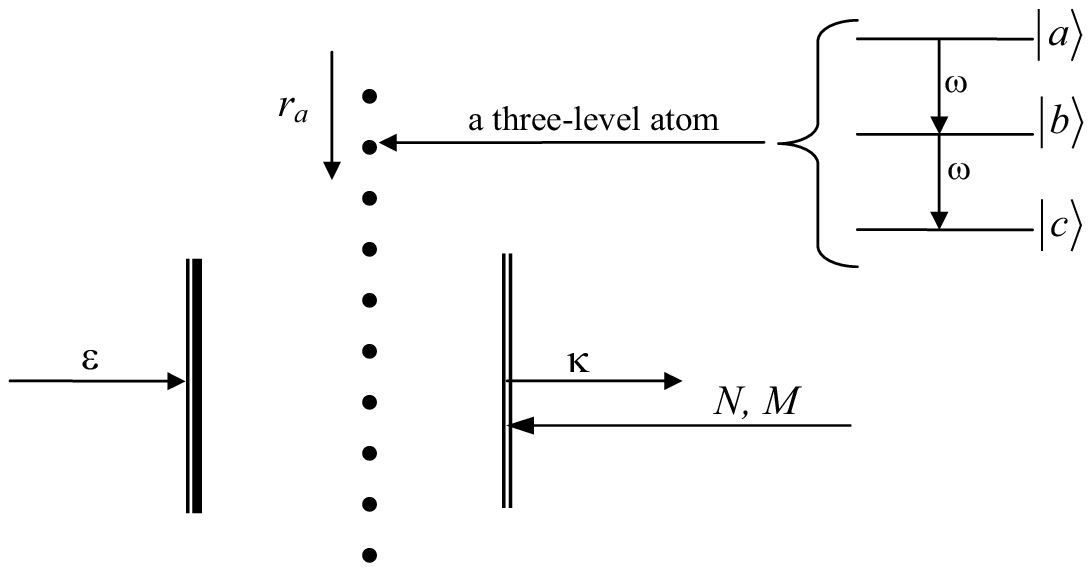}} {Fig.
2.1 {\footnotesize Schematic representation of the system under consideration .}}
\end{figure}
\end{center}
We take the initial state of a three-level atom to be
$$|\psi(0)\rangle=C_a(0)|a\rangle +C_c(0)|c\rangle\eqno(2.5)$$and
hence the initial density operator for a single atom has the form
$$\hat{\rho}_A=\rho_{aa}^{(0)}|a\rangle\langle a|+\rho_{ac}^{(0)}|a\rangle\langle
c|+\rho_{ca}^{(0)}|c\rangle\langle
a|+\rho_{cc}^{(0)}|c\rangle\langle
c|,\eqno(2.6a)$$where$$\rho_{aa}^{(0)}=|C_a|^2,\eqno(2.6b)$$
$$\rho_{ac}^{(0)}=C_a C_c^*,\eqno(2.6c)$$ $$\rho_{ca}^{(0)}=C_c C_a
^*,\eqno(2.6d)$$and$$\rho_{cc}^{(0)}=|C_c|^2.\eqno(2.6e)$$~With
the assumption that~$\rho_{ac}^{(0)}=\rho_{ca}^{(0)},$~the
equation of evolution of the density operator for the cavity mode,
commonly known as the master equation, is expressible as
~\cite{10}
$${d\over
dt}\hat\rho=p(2\hat{a}^\dag\hat{\rho}\hat{a}-\hat{\rho}\hat{a}\hat{a}^\dag-\hat{a}\hat{a}^\dag\hat{\rho})$$
$$~~~~+q(2\hat{a}\hat{\rho}\hat{a}^\dag-\hat{a}^\dag\hat{a}\hat{\rho}-\hat{\rho}\hat{a}^\dag\hat{a})$$
$$~~~~~~~~~~~~~~+u(\hat{a}^\dag\hat{\rho}\hat{a}^\dag-\hat{\rho}\hat{a}^{\dag
2})+v(\hat{a}^\dag\hat{\rho}\hat{a}^\dag-\hat{a}^{\dag
2}\hat{\rho})$$
$$~~~~~~~~~~~~+u^*(\hat{a}\hat{\rho}\hat{a}-\hat{a}^{2}\hat{\rho})+v^*(\hat{a}\hat{\rho}\hat{a}-\hat{\rho}\hat{a}^{2}),\eqno(2.7)$$ where
$$p={A\over 2B}\left[\rho_{aa}^{(0)}\left(1+{\Omega^2\over
4\gamma^2}\right)+\rho_{cc}^{(0)}\left({3\Omega^2\over
4\gamma^2}\right)-\rho_{ac}^{(0)}\left({3\Omega\over
2\gamma}\right)+{\kappa B\over A}N\right],\eqno(2.8a)$$
$$~~~~~~~q={A\over 2B}\left[\rho_{aa}^{(0)}\left({3\Omega^2\over
4\gamma^2}\right)+\rho_{cc}^{(0)}\left(1+{\Omega^2\over
4\gamma^2}\right)+\rho_{ac}^{(0)}\left({3\Omega\over
2\gamma}\right)+{\kappa B\over A}(N+1)\right],\eqno(2.8b)$$
$$u={A\over 2B}\bigg[-\rho_{aa}^{(0)}\left({\Omega\over
2\gamma}\right)\left(1-{\Omega^2\over
2\gamma^2}\right)+\rho_{cc}^{(0)}\left({\Omega\over
\gamma}\right)\left(1+{\Omega^2\over 4\gamma^2}\right)$$
$$-\rho_{ac}^{(0)}\left(1-{\Omega^2\over
2\gamma^2}\right)-{\kappa B\over A}M\bigg],\eqno(2.8c)$$
$$v={A\over 2B}\bigg[-\rho_{aa}^{(0)}\left({\Omega\over
\gamma}\right)\left(1+{\Omega^2\over
4\gamma^2}\right)+\rho_{cc}^{(0)}\left({\Omega\over
2\gamma}\right)\left(1-{\Omega^2\over 2\gamma^2}\right)$$
$$-\rho_{ac}^{(0)}\left(1-{\Omega^2\over
2\gamma^2}\right)-{\kappa B\over A}M\bigg],\eqno(2.8d)$$in which
$$A={2g^2r_a\over\gamma^2}\eqno(2.8e)$$is called the linear gain
coefficient,
$$N=\sinh^2 r,\eqno(2.8f)$$and
$$M=\cosh r\sinh r\eqno(2.8g)$$are reservoir parameters .
It is very important to note that the presence of the quadratic
terms~$\hat{a}^2$~and~$\hat{a}^{\dag 2}$~in the master equation is
a signature that the system under consideration may generate
squeezed light.

The expectation value of an operator~$\hat{A}$~evolves in time in
the Schr$\ddot{o}$dinger picture according to
$${d\over dt}\langle \hat{A}\rangle=Tr\left({d\hat\rho\over
dt}\hat{A}\right).\eqno(2.9)$$ With the aid of (2.7) and
(2.9), and employing the cyclic property of the trace operation
along with the relations
$$[\hat a, f(\hat a,\hat a^\dag)]={\partial\over\partial
\hat{a}^\dag}f(\hat a,\hat a^\dag),\eqno(2.10a)$$
$$[\hat a^\dag, f(\hat a,\hat a^\dag)]=-{\partial\over\partial
\hat{a}}f(\hat a,\hat a^\dag),\eqno(2.10b)$$it can easily be
verified that
$${d\over dt}\langle \hat{a}\rangle=(p-q)\langle \hat{a}\rangle+(u-v)\langle
\hat{a}^\dag\rangle,\eqno(2.11)$$
$${d\over dt}\langle \hat{a}^2\rangle=2[(p-q)\langle \hat{a}^2\rangle+(u-v)\langle
\hat{a}^\dag\hat{a}\rangle-v],\eqno(2.12)$$
$${d\over dt}\langle \hat{a}^\dag\hat{a}\rangle=2(p-q)\langle \hat{a}^\dag\hat{a}\rangle+(u-v)[\langle
\hat{a}^{\dag 2}\rangle+\langle
\hat{a}^{2}\rangle]+2p.\eqno(2.13)$$It proves to be convenient to
work with c-number variables than with operators. In view of this,
we wish to convert the operator equations into c-number equation
associated with the normal ordering. We note Eqs. (2.12), (2.13),
are already in the normal order. Hence one can write
$${d\over
dt}\langle\alpha(t)\rangle=-C\langle\alpha(t)\rangle+D\langle\alpha^*(t)\rangle,\eqno(2.14)$$
$${d\over
dt}\langle\alpha^2(t)\rangle=2[-C\langle\alpha^2(t)\rangle+D\langle\alpha^*(t)\alpha(t)\rangle-v],\eqno(2.15)$$
$${d\over dt}\langle \alpha^*(t)\alpha(t)\rangle=-2C\langle \alpha^*(t)\alpha(t)\rangle+D[\langle
\alpha^2(t)\rangle+\langle
\alpha^{*2}(t)\rangle]+2p,\eqno(2.16)$$where$$C=q-p,\eqno(2.17a)$$and$$D=u-v.\eqno(2.17b)$$Here~$\alpha$~and~$\alpha^*$~are
the c-number variables corresponding to the
operators~$\hat{a}$~and~$\hat{a}^\dag,$~respectively.

On the basis of (2.14), one can write the stochastic
differential equation
$${d\over
dt}\alpha(t)=-C\alpha(t)+D\alpha^*(t)+f(t),\eqno(2.18)$$where~$f(t)$~is
a noise force whose properties remain to be determined. We notice
that
$${d\over dt}\langle\alpha(t)\rangle=-C\langle\alpha(t)\rangle+D\langle\alpha^*(t)\rangle+\langle
f(t)\rangle.\eqno(2.19)$$ On comparing Eqs. (2.14) and (2.19), we
see that
$$\langle f(t)\rangle=0.\eqno(2.20)$$We note that
$${d\over
dt}\langle\alpha^2(t)\rangle=2\langle\alpha(t)\dot{\alpha}(t)\rangle.\eqno(2.21)$$On
combining (2.21) with (2.18), there follows
$${d\over
dt}\langle\alpha^2(t)\rangle=2[-C\langle\alpha^2(t)\rangle+D\langle\alpha^*(t)\alpha(t)\rangle+\langle\alpha(t)f(t)\rangle].\eqno(2.22)$$Comparison
of Eqs. (2.15) and (2.22) shows that
$$\langle\alpha(t)f(t)\rangle=-v.\eqno(2.23)$$We also note that
$${d\over dt}\langle
\alpha^*(t)\alpha(t)\rangle=\langle\alpha^*(t)\dot{\alpha}(t)\rangle+\langle\alpha(t)\dot{\alpha}^*(t)\rangle\eqno(2.24)$$
Employing Eq. (2.18) and its complex conjugate, we arrive at
$${d\over dt}\langle \alpha^*(t)\alpha(t)\rangle=-2C\langle \alpha^*(t)\alpha(t)\rangle+D[\langle
\alpha^2(t)\rangle+\langle \alpha^{*2}(t)\rangle]$$
$$+\langle
\alpha(t)f^*(t)\rangle+\langle
\alpha^*(t)f(t)\rangle.\eqno(2.25)$$ Comparison of this with Eq.
(2.16) indicates that
$$\langle\alpha(t)f^*(t)\rangle+\langle\alpha^*(t)f(t)\rangle=2p.\eqno(2.26)$$Furthermore,  a formal solution of (2.18) can be written as
$$\alpha(t)=\alpha(0)e^{-ct}+\int_0^t~e^{-c(t-t')}~[D\alpha^*(t')+f(t')]dt'.\eqno(2.27)$$Multiplying (2.27) by ~$f(t)$~and taking the expectation value of both sides, we get
$$\langle\alpha(t)f(t)\rangle=\langle\alpha(0)f(t)\rangle e^{-ct}+\int_0^t~e^{-c(t-t')}~[D\langle\alpha^*(t')f(t)\rangle+\langle f(t')f(t)\rangle]dt'.\eqno(2.28)$$With the assumption that the noise force at a time t and a cavity mode variable at an earlier time~$t'$~ are not correlated, we can set
$$\langle \alpha(t')f(t)\rangle=\langle \alpha(t')\rangle \langle f(t)\rangle.\eqno(2.29)$$Thus on account of (2.20), we have
$$\langle \alpha(t')f(t)\rangle=0.\eqno(2.30)$$Hence in view of (2.23) and (2.30), Eq.(2.28) reduces to
$$\int_0^t~e^{-c(t-t')}\langle f(t')f(t)\rangle dt'=-v.\eqno(2.31)$$On the basis of this result, one can write
$$\langle f(t')f(t)\rangle=-2v\delta(t-t').\eqno(2.32)$$Moreover, one easily obtains
$$\langle\alpha(t)f^*(t)\rangle+\langle\alpha^*(t)f(t)\rangle=\langle\alpha(0)f^*(t)\rangle e^{-ct}+\langle\alpha^*(0)f(t)\rangle e^{-ct}$$
$$+\int_0^t~e^{-c(t-t')}~[D\langle\alpha^*(t')f^*(t)\rangle+\langle f(t')f^*(t)\rangle]dt'$$
$$+\int_0^t~e^{-c(t-t')}~[D\langle\alpha(t')f(t)\rangle+\langle f^*(t')f(t)\rangle]dt'.\eqno(2.33)$$Applying (2.30) and (2.26), we get
$$\int_0^t~e^{-c(t-t')}~[\langle f(t')f^*(t)\rangle+\langle f^*(t')f(t)\rangle]dt'=2p.\eqno(2.34)$$On the basis of this result, one can write
$$\langle f(t)f^*(t')\rangle=2p\delta(t-t').\eqno(2.35)$$It is worth mentioning that Eqs. (2.20), (2.32) and (2.35) describe the correlation properties of the noise force ~$f(t)$~ associated with the normal ordering.

Now introducing a new variable defined by
$$\alpha_{\pm}(t)=\alpha^*(t)\pm\alpha(t),\eqno(2.36)$$we easily obtain
$${d\over dt}\alpha_{\pm}(t)=-\lambda_{\mp}\alpha_{\pm}(t)+f^*(t)\pm f(t),\eqno(2.37a)$$where
$$\lambda_{\mp}(t)=C\mp D.\eqno(2.37b)$$
The solution of Eq. (2.37a) can be expressed as
$$\alpha_{\pm}(t)=\alpha_{\pm}(0)e^{-\lambda_{\mp}(t)}+\int_0^t
e^{-\lambda_{\mp}(t-t')}[f^*(t')\pm f(t')]dt'.\eqno(2.38)$$It then
follows that
$$\alpha(t)=A(t)\alpha(0)+B(t)\alpha^*(0)+F(t),\eqno(2.39)$$
where
$$A(t)={1\over 2}\left(e^{-\lambda_-t}+e^{-\lambda_+t}\right),\eqno(2.40a)$$
$$B(t)={1\over
2}\left(e^{-\lambda_-t}-e^{-\lambda_+t}\right),\eqno(2.40b)$$and
$$F(t)=F_+(t)+F_-(t),\eqno(2.40c)$$with
$$F_{\pm}(t)={1\over 2}\int_0^t e^{-\lambda_{\mp}(t-t')}[f(t')\pm f^*(t')]dt'.\eqno(2.40d)$$
\section{Quadrature Fluctuations}
\label{sec:results}
In this section we wish to calculate the
quadrature variance and squeezing spectrum for the cavity mode
under consideration.
\subsection{Quadrature Variance}
Here we seek to
calculate the variance of the quadrature operators
$$ \hat{a}_+=\hat{a}^\dag +\hat{a},\eqno(3.1a)$$and
$$ \hat{a}_-=i(\hat{a}^\dag -\hat{a}).\eqno(3.1b)$$With the aid of (3.1a) and
(3.1b), one can express the quadrature variance in the normal
order as
$$\Delta a^2_{\pm}=1\pm\langle(\hat{a}^{\dag 2}+\hat{a}^2\pm
2\hat{a}^{\dag}\hat{a})\rangle\mp\langle(\hat{a}^{\dag
}\pm\hat{a})\rangle^2\eqno(3.2)$$and the c-number expression
corresponding to (3.2) is
$$\Delta a^2_{\pm}=1\pm\langle \alpha^2_{\pm}(t)\rangle\mp\langle
\alpha_{\pm}(t)\rangle^2,\eqno(3.3)$$where~$\alpha_{\pm}(t)$~is
defined by Eq. (2.36). We consider here the case for which the
cavity mode is initially in a vacuum state. Hence on account of
(2.20) and (2.38), we see that
$$\langle\alpha_{\pm}(t)\rangle=0.\eqno(3.4)$$In view of this
result, Eq.(3.3) reduces to
$$\Delta a^2_{\pm}=1\pm\langle\alpha^2_{\pm}(t)\rangle.\eqno(3.5)$$Moreover,
employing Eq. (2.37a), one easily gets
$${d\over dt}\langle\alpha^2_{\pm}(t)\rangle=2[-\lambda_{\mp}\langle\alpha^2_{\pm}(t)\rangle+\langle\alpha_{\pm}(t)f^*(t)\rangle \pm \langle\alpha_{\pm}(t)f(t)\rangle
].\eqno(3.6)$$Multiplying Eq. (2.36) by~$f(t)$~and taking into
account
 Eq. (2.23) and Eq. (2.26), we have
$$\langle\alpha_{\pm}(t)f(t)\rangle=p\mp v,\eqno(3.7a)$$similarly,
one readily finds
$$\langle\alpha_{\pm}(t)f^*(t)\rangle=-v\pm
p.\eqno(3.7b)$$Therefore, in view of this result, Eq. (3.6) can be
rewritten as
$${d\over dt}\langle\alpha^2_{\pm}(t)\rangle=2[-\lambda_{\mp}\langle\alpha^2_{\pm}(t)\rangle-2v\pm
2p].\eqno(3.8)$$With the cavity mode initially in a vacuum state,
the solution of this equation has the form
$$\langle\alpha^2_{\pm}(t)\rangle=\left({-2v\pm
2p\over\lambda_{\mp}}\right)(1-e^{-2\lambda_{\mp}t}).\eqno(3.9)$$Now
 combination of Eqs. (3.5) and (3.9) yields
$$\Delta a^2_{\pm}=1\pm\left({-2v\pm
2p\over\lambda_{\mp}}\right)(1-e^{-2\lambda_{\mp}t})\eqno(3.10)$$and
at steady state this equation reduces to
$$\Delta a^2_{\pm}={\lambda_{\mp}+2p\mp
2v\over\lambda_{\mp}}.\eqno(3.11)$$ Upon substituting Eq. (2.37b)
along with  Eqs. (2.17a) and (2.17b) into Eq. (3.11), one arrives
at the result
$$\Delta a^2_{\pm}={(q\mp u)+(p\mp v)\over(q\mp u)-(p\mp
v)},\eqno(3.12)$$ where
$$(q\mp u)={A\over 2B}\bigg[{\kappa B\over A}(1+N\pm
M)+\rho_{aa}^{(0)}\left({3\Omega^2\over 4\gamma^2}\pm{\Omega\over
 2\gamma}\left(1-{\Omega^2\over 2\gamma^2}\right)\right)$$
$$+\rho_{cc}^{(0)}\left(\left(1+{\Omega^2\over
4\gamma^2}\right)\mp{\Omega\over \gamma}\left(1+{\Omega^2\over
4\gamma^2}\right)\right)+\rho_{ac}^{(0)}\left({3\Omega\over
2\gamma}\pm\left(1-{\Omega^2\over
2\gamma^2}\right)\right)\bigg],\eqno(3.13a)$$and
$$(p\mp v)={A\over 2B}\bigg[{\kappa B\over A}(N\pm
M)+\rho_{aa}^{(0)}\left(\left(1+{\Omega^2\over
4\gamma^2}\right)\pm{\Omega\over \gamma}\left(1+{\Omega^2\over
4\gamma^2}\right)\right)$$ $$+\rho_{cc}^{(0)}\left({3\Omega^2\over
4\gamma^2}\mp{\Omega\over 2\gamma}\left(1-{\Omega^2\over
2\gamma^2}\right)\right) +\rho_{ac}^{(0)}\left(-{3\Omega\over
2\gamma}\pm\left(1-{\Omega^2\over
2\gamma^2}\right)\right)\bigg].\eqno(3.13b)$$

It proves to be more convenient to introduce a new parameter
defined by
$$\rho_{aa}^{(0)}={1-\eta\over 2},\eqno(3.14a)$$so that in view of
the fact that
$$\rho_{aa}^{(0)}+\rho_{cc}^{(0)}=1,\eqno(3.14b)$$we have
$$\rho_{cc}^{(0)}={1+\eta\over 2}.\eqno(3.14c)$$Using the
relations
$$\rho_{ac}^{(0)}=|\rho_{ac}^{(0)}|e^{i\theta},\eqno(3.15a)$$
$$\rho_{ca}^{(0)}=|\rho_{ac}^{(0)}|e^{-i\theta},\eqno(3.15b)$$and
 the assumption that~$\rho_{ac}^{(0)}=\rho_{ca}^{(0)},$~one easily
obtains
$$|\rho_{ac}^{(0)}|^2=\rho_{ac}^{(0)}\rho_{ca}^{(0)}.\eqno(3.16a)$$Now
on account of Eqs. (2.6b), (2.6c), (2.6d) and (2.6e), we arrive at
$$\rho_{ac}^{(0)}\rho_{ca}^{(0)}=\rho_{aa}^{(0)}\rho_{cc}^{(0)},\eqno(3.16b)$$
so that in view of Eq. (3.16a) and (3.16b), we have
$$|\rho_{ac}^{(0)}|^2=\rho_{aa}^{(0)}\rho_{cc}^{(0)}.\eqno(3.17a)$$Hence with
the aid of Eq. (3.14a) and (3.14c) Eq. (3.17a) takes the form
$$\rho_{ac}^{(0)}={1\over 2}(1-\eta^2)^{1/2}.\eqno(3.17b)$$Now on
combining Eqs.  (3.13a), (3.13b),(3.14a), (3.14c), and (3.17b),
there follows
$$(q\mp u)+(p\mp v)=\bigg[{\kappa\over 2}(1+2N\pm 2M)+{A(\beta(2\beta\mp
3\eta)\mp(1-\eta^2)^{1\over 2}(\beta^2-2)+2)\over
2(1+\beta^2)(1+{\beta^2\over 4})} \bigg]\eqno(3.18a)$$and
$$(q\mp u)-(p\mp v)=\bigg[{\kappa\over 2}+{A(\eta(2-\beta^2)\mp\beta(1+\beta^2)+3\beta(1-\eta^2)^{1\over
2})\over 2(1+\beta^2)(1+{\beta^2\over 4})}
\bigg],\eqno(3.18b)$$where
$$B=(1+{\Omega^2\over\gamma^2})(1+{\Omega^2\over4\gamma^2})\eqno(3.19a)$$and
$$\beta={\Omega\over\gamma}.\eqno(3.19b)$$ In view of (3.18), the
variance of the quadrature operators takes the form
$$\Delta a^2_{\pm}={2\kappa(1+\beta^2)(1+{\beta^2\over 4}) e^{\pm 2r}+A\left[\beta(2\beta\mp 3\eta)\mp(1-\eta^2)^{1\over
2}(\beta^2-2)+2\right]\over2\kappa(1+\beta^2)(1+{\beta^2\over
4})+A\left[\eta(2-\beta^2)\mp\beta(1+\beta^2)+3\beta(1-\eta^2)^{1\over
2}\right]},\eqno(3.20)$$in which we have made use of Eqs. (2.8f)
and (2.8g).

 We next seek  to consider some special cases of interest.
We then note that \\for~$\beta=0,$~expression (3.20) reduces to
$$\Delta a_{+}^2={\kappa e^{2r}+A[1+(1-\eta^2)^{1\over 2}]\over(\eta A+\kappa)},\eqno(3.21)$$and
$$\Delta a_{-}^2={\kappa e^{-2r}+A[1-(1-\eta^2)^{1\over 2}]\over(\eta
A+\kappa)}.\eqno(3.22)$$This represents the quadrature variance in
the absence of the driving light.
\begin{center}
\begin{figure}[h]
\centerline{\includegraphics [width=2.5 in]{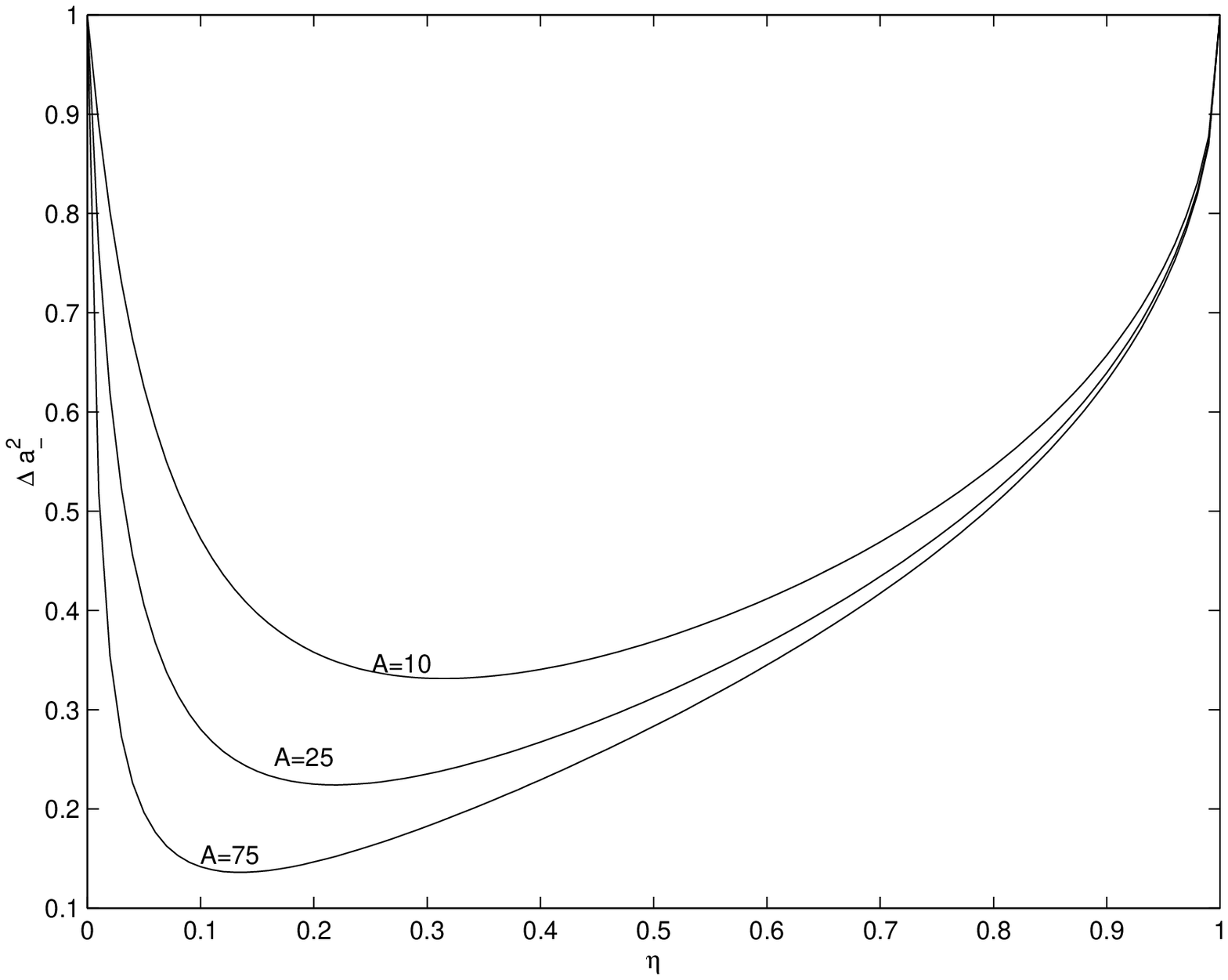}} {Fig.
3.1 {\footnotesize Plots of the quadrature variance~$\Delta
a^2_-$~versus~$\eta$~for~$r=0,$~ $\kappa=0.8,$  ~$\beta =0,$~and
for different values of the linear gain coefficient .}}
\end{figure}
\end{center}
\begin{center}
\begin{figure}[h]
\centerline{\includegraphics [width=2.5 in]{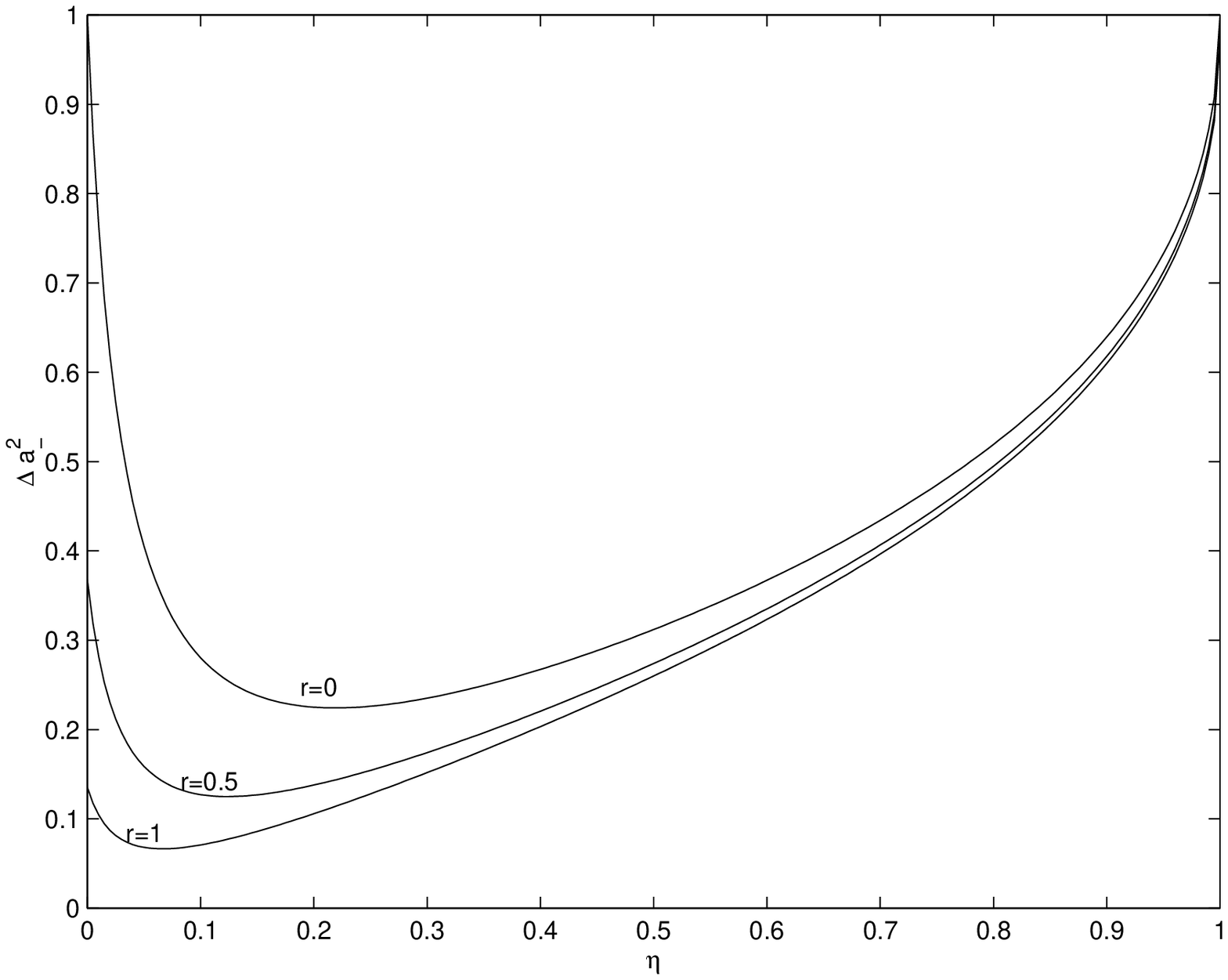}} {Fig.
3.2 {\footnotesize Plots of the quadrature variance~$\Delta
a^2_-$~versus~$\eta$~for~$A=25,$~ $\kappa=0.8,$  ~$\beta =0$~and
for different values of the squeeze parameter .}}
\end{figure}
\end{center}
Fig. (3.1) or (3.2) shows that the cavity mode is in a squeezed
state for all values of~$\eta$~between zero and one and the degree
of squeezing increases with the linear gain coefficient~$A$~or the
squeeze parameter~$r.$~
Moreover, we want to consider the case when all atoms are
initially in the upper level and when the upper and lower levels
are coupled by a strong coherent light. Thus upon setting
~$\eta=-1$~in Eq. (3.20), we have
$$\Delta a^2_{+}={2\kappa(1+\beta^2)(1+{\beta^2\over 4}) e^{ 2r}+A\left[\beta(2\beta+ 3)+2\right]\over2\kappa(1+\beta^2)(1+{\beta^2\over
4})+A\left[(\beta^2-2)-\beta(1+\beta^2)\right]}\eqno(3.23a)$$and
$$\Delta a^2_{-}={2\kappa(1+\beta^2)(1+{\beta^2\over 4}) e^{- 2r}+A\left[\beta(2\beta- 3)+2\right]\over2\kappa(1+\beta^2)(1+{\beta^2\over
4})+A\left[(\beta^2-2)+\beta(1+\beta^2)\right]}.\eqno(3.23b)$$
\begin{center}
\begin{figure}[h]
\centerline{\includegraphics [width=2.5 in]{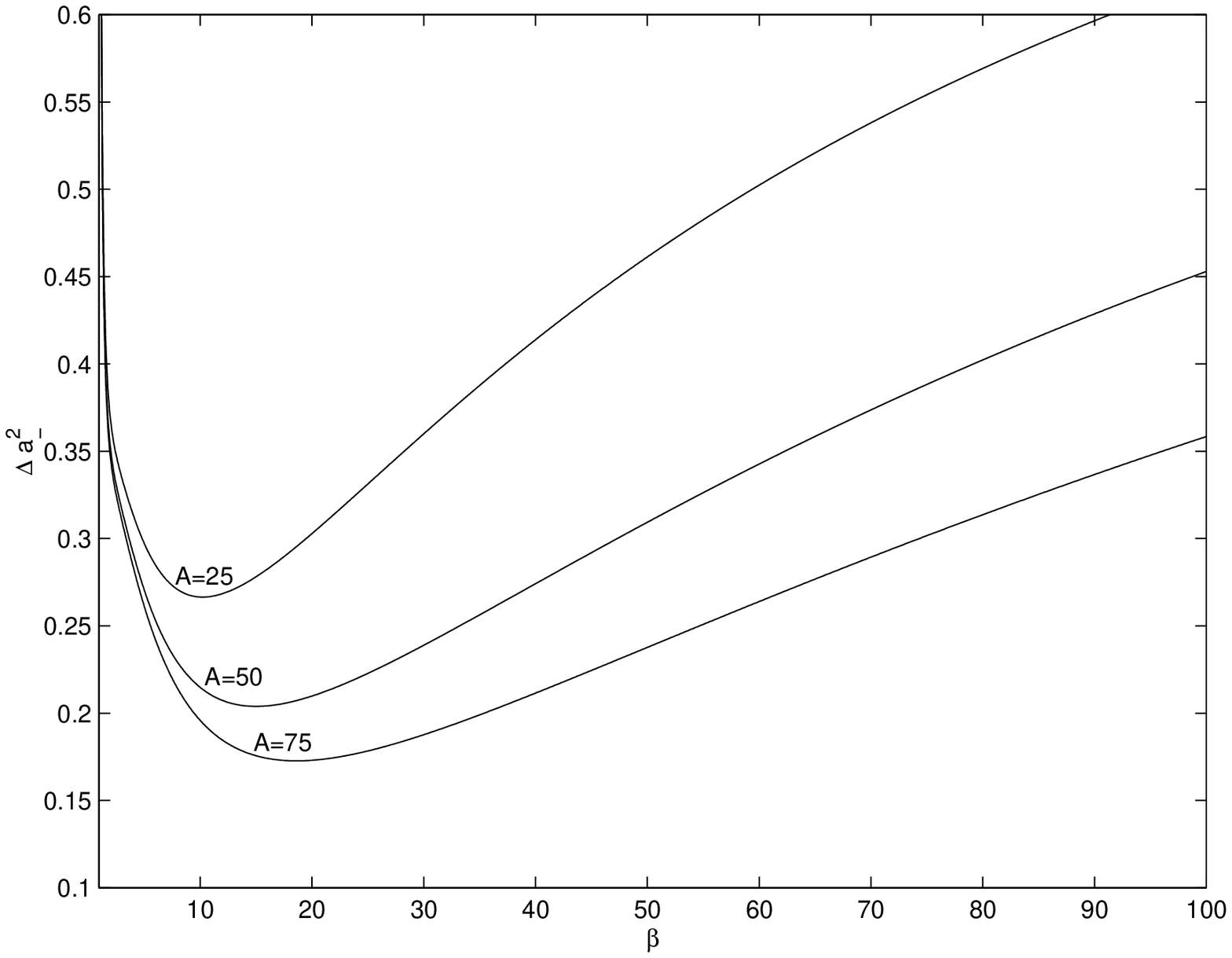}} {Fig.
3.3 {\footnotesize Plots of the quadrature variance~$\Delta
a^2_-$~versus~$\beta$~for~$r=0,$~ $\kappa=0.8,$  ~$\eta =-1,$~and
for different values of the linear gain coefficient .}}
\end{figure}
\end{center}

 From Fig. (3.3) we note that a relatively better squeezing can
be achieved for large values of the linear gain coefficient
~$A.$~And Fig. (3.4) indicates that the degree of squeezing
increases with the squeeze parameter~$r.$~
\begin{center}
\begin{figure}[h]
\centerline{\includegraphics [width=2.5 in]{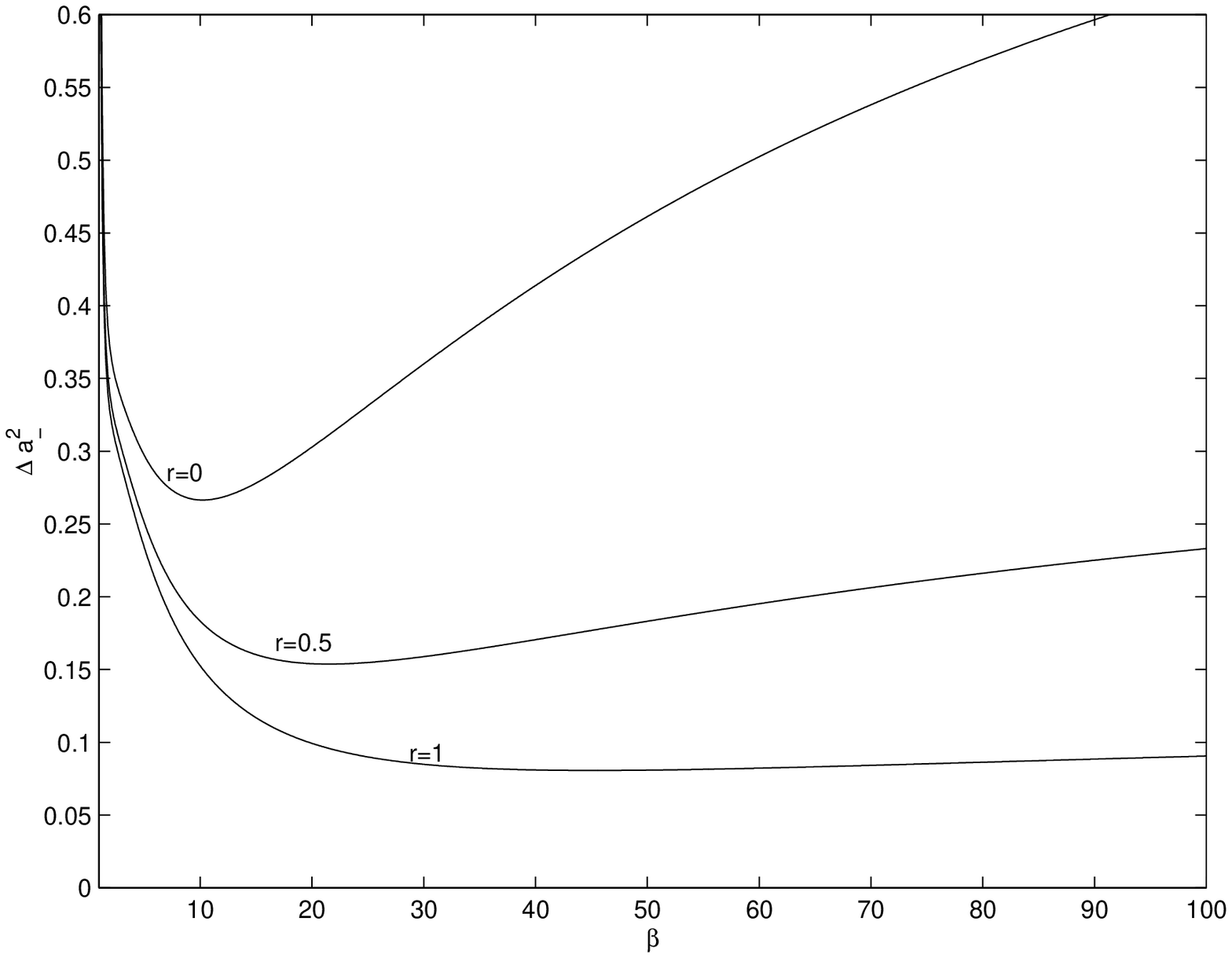}} {Fig.
3.4 {\footnotesize Plots of the quadrature variance~$\Delta
a^2_-$~versus~$\beta$~for~$A=25,$~ $\kappa=0.8,$  ~$\eta =-1,$~and
for different values of the squeeze parameter .}}
\end{figure}
\end{center}
 Furthermore, we want
to consider the case when half of the atoms are initially in the
upper level while the remaining half are in the lower level and
when the upper and the lower levels are coupled by a strong
coherent light. Thus upon setting~$\eta=0$~in Eq. (3.20), we have

$$\Delta a^2_{+}={2\kappa(1+\beta^2)(1+{\beta^2\over
4}) e^{ 2r}+A\beta^2\over2\kappa(1+\beta^2)(1+{\beta^2\over
4})+A\left[\beta(2-\beta^2)\right]}\eqno(3.24a)$$and
$$\Delta a^2_{-}={2\kappa (1+\beta^2)(1+{\beta^2\over 4})e^{- 2r}+3A\beta^2\over2\kappa(1+\beta^2)(1+{\beta^2\over
4})+A\left[\beta(4+\beta^2)\right]}.\eqno(3.24b)$$

\begin{center}
\begin{figure}[h]
\centerline{\includegraphics [width=2.5 in]{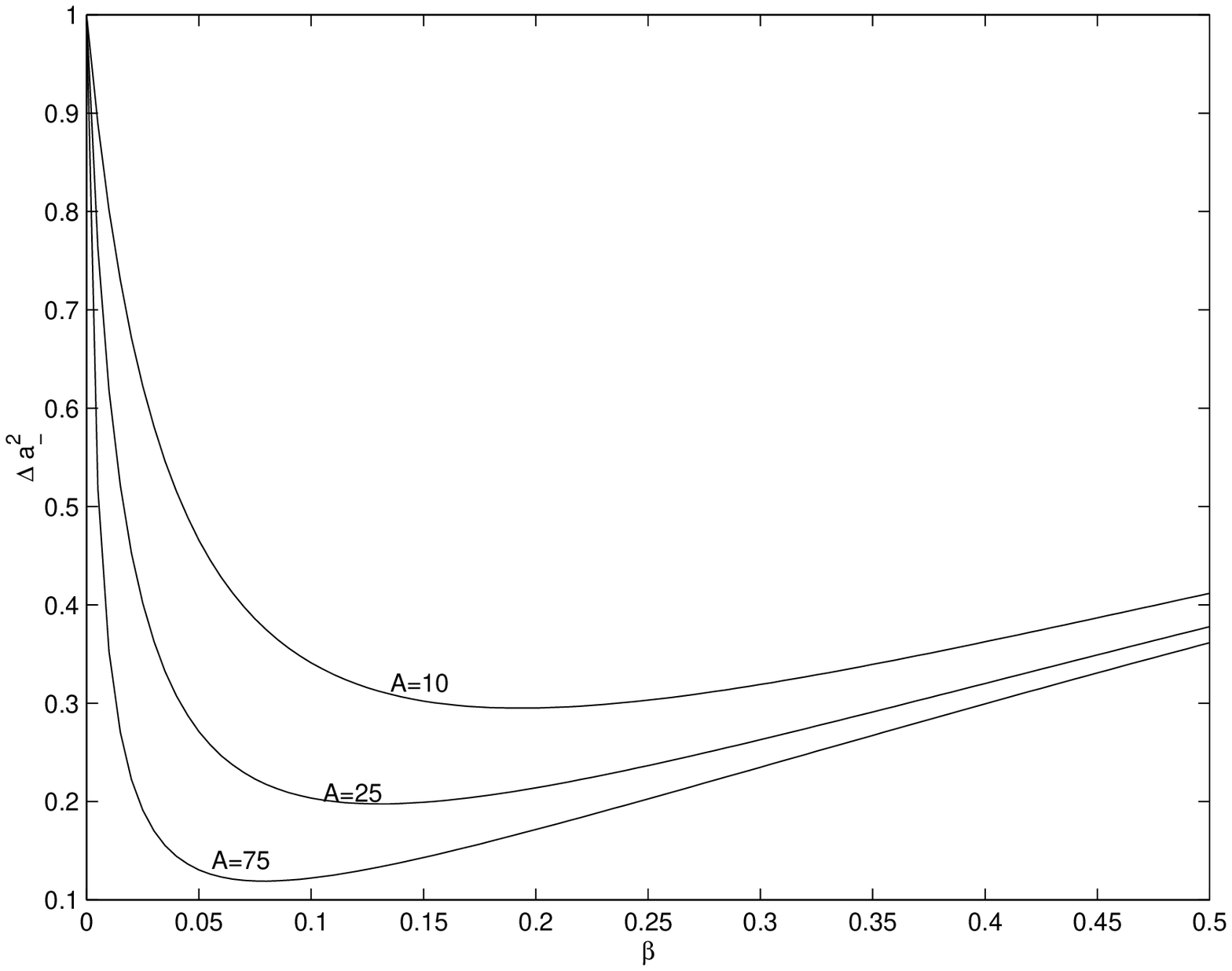}} {Fig.
3.5 {\footnotesize Plots of the quadrature variance~$\Delta
a^2_-$~versus~$\beta$~for~$r=0,$~ $\kappa=0.8,$  ~$\eta =0,$~and
for different values of the linear gain coefficient .}}
\end{figure}
\end{center}

 Fig. (3.5) or (3.6) indicates  that the degree of squeezing
increases with the linear gain coefficient ~$A$~ or the squeeze
parameter~$r.$~We also note that a relatively better squeezing can
be achieved  for very small values of ~$\beta.$~
\begin{center}
\begin{figure}[h]
\centerline{\includegraphics [width=2.5 in]{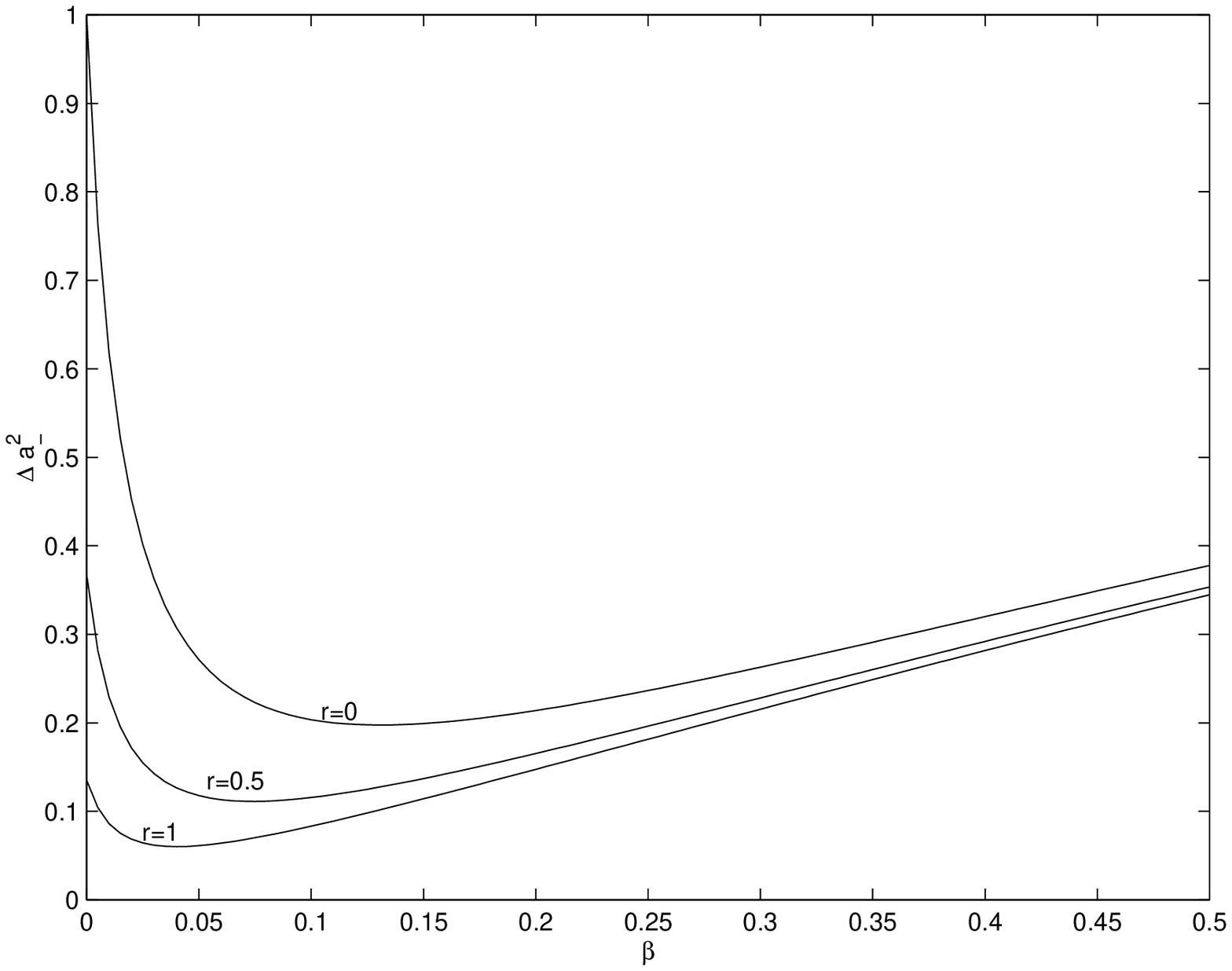}} {Fig.
3.6 {\footnotesize Plots of the quadrature variance~$\Delta
a^2_-$~versus~$\beta$~for~$A=25,$~ $\kappa=0.8,$  ~$\eta =0,$~and
for different values of the squeeze parameter .}}
\end{figure}
\end{center}
\subsection{Squeezing Spectrum}
\label{sec:logicmachine}
The squeezing spectrum of the output radiation is defined by
~\cite{2}.
$$S_{\pm}^{out}(\omega)=2Re\int_0^{\infty}\langle
\hat{a}_{\pm}^{out}(t),\hat{a}_{\pm}^{out}(t+\tau)\rangle_{ss}~e^{i\omega\tau}
d\tau,\eqno(3.25)$$where ~$ss$~stands for steady state and
$$\hat{a}_{+}^{out}(t)=\hat{a}_{out}^\dag(t)+\hat{a}_{out}(t),\eqno(3.26a)$$
$$\hat{a}_{-}^{out}(t)=i(\hat{a}_{out}^\dag(t)-\hat{a}_{out}(t)).\eqno(3.26b)$$Making
use of Eqs. (3.26a) and (3.26b) along with
$$\langle\hat{A},\hat{B}\rangle=\langle\hat{A}\hat{B}\rangle-\langle\hat{A}\rangle\langle\hat{B}\rangle,\eqno(3.27)$$ one
finds
$$\langle\hat{a}_{\pm}^{out}(t),\hat{a}_{\pm}^{out}(t+\tau)\rangle=\langle\hat{a}_{out}^\dag(t)\hat{a}_{out}(t+\tau)\rangle+\langle\hat{a}_{out}(t)\hat{a}_{out}^\dag(t+\tau)\rangle$$
$$\pm\langle\hat{a}_{out}^\dag(t)\hat{a}_{out}^\dag(t+\tau)\rangle\pm\langle\hat{a}_{out}(t)\hat{a}_{out}(t+\tau)\rangle$$
$$\mp\langle\hat{a}_{out}^\dag(t)\pm\hat{a}_{out}(t)\rangle\langle\hat{a}_{out}^\dag(t+\tau)\pm\hat{a}_{out}(t+\tau)\rangle.\eqno(3.28)$$
With the aid of
$$[\hat{a}(t),\hat{a}^\dag(t+\tau)]=\delta(\tau),\eqno(3.29)$$
one can put Eq. (3.28) in the normal order as
$$\langle\hat{a}_{\pm}^{out}(t),\hat{a}_{\pm}^{out}(t+\tau)\rangle=\delta(\tau)+\langle\hat{a}_{out}^\dag(t)\hat{a}_{out}(t+\tau)\rangle+\langle\hat{a}_{out}^\dag(t+\tau)\hat{a}_{out}(t)\rangle$$
$$\pm\langle\hat{a}_{out}^\dag(t)\hat{a}_{out}^\dag(t+\tau)\rangle+\langle\hat{a}_{out}(t)\hat{a}_{out}(t+\tau)\rangle$$
$$\mp\langle\hat{a}_{out}^\dag(t)\pm\hat{a}_{out}(t)\rangle\langle\hat{a}_{out}^\dag(t+\tau)\pm\hat{a}_{out}(t+\tau)\rangle.\eqno(3.30)$$ The c-number equation corresponding to Eq. (3.30)
is
$$\langle\hat{a}_{\pm}^{out}(t),\hat{a}_{\pm}^{out}(t+\tau)\rangle=\delta(\tau)\pm\langle\alpha_{out}^*(t)\alpha_{out}^*(t+\tau)\rangle+\langle\alpha_{out}^*(t)\alpha_{out}(t+\tau)\rangle$$
$$+\langle\alpha_{out}^*(t+\tau)\alpha_{out}(t)\rangle\pm\langle\alpha_{out}(t)\alpha_{out}(t+\tau)\rangle$$
$$\mp\langle\alpha_{out}^*(t)\pm\alpha_{out}(t)\rangle\langle\alpha_{out}^*(t+\tau)\pm\alpha_{out}(t+\tau)\rangle.\eqno(3.31)$$With
the aid of (3.27) this can be rewritten as
$$\langle\hat{a}_{\pm}^{out}(t),\hat{a}_{\pm}^{out}(t+\tau)\rangle=\delta(\tau)\pm\langle\alpha_{\pm}^{out}(t),\alpha_{\pm}^{out}(t+\tau)\rangle,\eqno(3.32)$$where
$$\alpha_{\pm}^{out}(t)=\alpha_{out}^*(t)\pm\alpha_{out}(t).\eqno(3.33)$$ Substitution
of (3.32) into Eq. (3.25) yields
$$S_{\pm}^{out}(\omega)=2Re\int_0^{\infty}\delta(\tau)~e^{i\omega\tau}d\tau\pm
2Re\int_0^{\infty}\langle\alpha_{\pm}^{out}(t),\alpha_{\pm}^{out}(t+\tau)\rangle_{ss}~e^{i\omega\tau}d\tau.\eqno(3.34)$$It
proves to be more convenient to rewrite the first integral in
(3.34) as
$$2Re\int_0^{\infty}\delta(\tau)~e^{i\omega\tau}d\tau=\int_0^{\infty}\delta(\tau)~e^{i\omega\tau}d\tau+\int_0^{\infty}\delta(\tau)~e^{-i\omega\tau}d\tau.\eqno(3.35)$$Upon
replacing ~$\tau$~by ~$-\tau$~in the second integral, one readily
finds
$$2Re\int_0^{\infty}\delta(\tau)~e^{i\omega\tau}d\tau=\int_{-\infty}^{\infty}\delta(\tau)~e^{i\omega\tau}d\tau=1.\eqno(3.36)$$On
account of this result, Eq. (3.34) takes the form
$$S_{\pm}^{out}(\omega)=1\pm2Re\int_0^{\infty}\langle\alpha_{\pm}^{out}(t),\alpha_{\pm}^{out}(t+\tau)\rangle_{ss}~e^{i\omega\tau}d\tau.\eqno(3.37)$$
For a cavity mode coupled to a squeezed vacuum reservoir, the
output and intracavity variables are related by
$$\alpha_{\pm}^{out}(t)=\sqrt\kappa\alpha_{\pm}(t)-\alpha_{in\pm}(t),\eqno(3.38)$$where
$$\alpha_{\pm}^{out}(t)=\alpha_{out}^*(t)\pm\alpha_{out}(t),\eqno(3.39a)$$and
$$\alpha_{in\pm}(t)={1\over\sqrt\kappa}(f_R^*(t)\pm
f_R(t)),\eqno(3.39b)$$with~$f_R(t)$~being the noise force
associated with the reservoir. Therefore on account of Eqs. (3.38)
and (3.4) the squeezing spectrum can be put in the form
$$S_{\pm}^{out}(\omega)=1\pm 2\kappa
Re\int_0^{\infty}\langle\alpha_{\pm}(t)\alpha_{\pm}(t+\tau)\rangle_{ss}~e^{i\omega\tau}d\tau$$
$$\mp2\sqrt\kappa Re\int_0^{\infty}\langle\alpha_{\pm}(t)\alpha_{in\pm}(t+\tau)\rangle_{ss}~e^{i\omega\tau}d\tau$$
$$\mp2\sqrt\kappa Re\int_0^{\infty}\langle\alpha_{in\pm}(t)\alpha_{\pm}(t+\tau)\rangle_{ss}~e^{i\omega\tau}d\tau$$
$$\pm 2
Re\int_0^{\infty}\langle\alpha_{in\pm}(t)\alpha_{in\pm}(t+\tau)\rangle_{ss}~e^{i\omega\tau}d\tau.\eqno(3.40)$$

Furthermore, the solution of Eq. (2.37a) can be written as
$$\alpha_{\pm}(t+\tau)=\alpha_{\pm}(t)~e^{-\lambda_{\mp}\tau}+e^{-\lambda_{\mp}\tau}\int_0^{\tau}e^{\lambda_{\mp}\tau'}[f^*(t+\tau')\pm
f(t+\tau')]d\tau'.\eqno(3.41)$$Multiplying Eq. (3.41) by
~$\alpha_{\pm}(t)$~and taking the expectation value of both sides,
we get
$$\langle\alpha_{\pm}(t)\alpha_{\pm}(t+\tau)\rangle=\langle\alpha_{\pm}^2(t)\rangle~e^{-\lambda_{\mp}\tau}$$
$$+e^{-\lambda_{\mp}\tau}\int_0^{\tau}e^{\lambda_{\mp}\tau'}[\langle\alpha_{\pm}(t)f^*(t+\tau')\rangle\pm
\langle\alpha_{\pm}(t)f(t+\tau')\rangle]d\tau'.\eqno(3.42)$$On
account of the fact that the reservoir noise force at time t and a
cavity mode variable at an earlier time are uncorrelated, one can
write
$$\langle\alpha_{\pm}(t)f^*(t+\tau')\rangle=\langle\alpha_{\pm}(t)\rangle\langle
f^*(t+\tau')\rangle=0,\eqno(3.43a)$$
$$\langle\alpha_{\pm}(t)f(t+\tau')\rangle=\langle\alpha_{\pm}(t)\rangle\langle
f(t+\tau')\rangle=0,\eqno(3.43b)$$where we have made use of (3.4).
In view of Eqs. (3.43), Eq. (3.42) takes the form
$$\langle\alpha_{\pm}(t)\alpha_{\pm}(t+\tau)\rangle=\langle\alpha_{\pm}^2(t)\rangle~e^{-\lambda_{\mp}\tau}.\eqno(3.44a)$$With
the aid of Eq. (3.9), Eq. (3.44a) becomes
$$\langle\alpha_{\pm}(t)\alpha_{\pm}(t+\tau)\rangle_{ss}=\left({-2v\pm
2p\over\lambda_{\mp}}\right)~e^{-\lambda_{\mp}\tau}.\eqno(3.44b)$$

Furthermore, multiplying Eq.(3.39b) by~$\alpha_{\pm}(t)$~and
taking the expectation value of both sides, we have
$$\langle\alpha_{\pm}(t)\alpha_{in\pm}(t+\tau)\rangle={1\over\sqrt\kappa}(\langle \alpha_{\pm}(t)f_R^*(t+\tau)\rangle\pm
\langle \alpha_{\pm}(t)f_R(t+\tau)\rangle),\eqno(3.45)$$in view of
the fact that the reservoir force at a time t and cavity mode at
an earlier time are uncorrelated, one can write
$$\langle\alpha_{\pm}(t)f_R^*(t+\tau)\rangle=\langle\alpha_{\pm}(t)\rangle\langle
f_R^*(t+\tau)\rangle,\eqno(3.46a)$$
$$\langle\alpha_{\pm}(t)f_R(t+\tau)\rangle=\langle\alpha_{\pm}(t)\rangle\langle
f_R(t+\tau)\rangle,\eqno(3.46b)$$ so that with the aid of Eqs.
(3.4), (3.46), Eq. (3.45) becomes
$$\langle\alpha_{\pm}(t)\alpha_{in\pm}(t+\tau)\rangle=0,\eqno(3.47)$$

Moreover, multiplying Eq. (3.41) by ~$\alpha_{in\pm}(t)$~and
taking the expectation value of both sides, we find
$$\langle\alpha_{in\pm}(t)\alpha_{\pm}(t+\tau)\rangle=\langle\alpha_{in\pm}(t)\alpha_{\pm}(t)\rangle~e^{-\lambda_{\mp}\tau}$$
$$+e^{-\lambda_{\mp}\tau}\int_0^\tau
e^{\lambda_{\mp}\tau'}[\langle\alpha_{in\pm}(t)f^*(t+\tau')\rangle\pm\langle\alpha_{in\pm}(t)f(t+\tau')\rangle]d\tau',\eqno(3.48)$$so
that with the help of (2.38) and (3.39b), we have
$$\langle\alpha_{in\pm}(t)\alpha_{\pm}(t+\tau)\rangle={e^{-\lambda_{\mp}\tau}\over\sqrt\kappa}\langle f_R^*(t)\alpha_{\pm}(0)\rangle
e^{-\lambda_{\mp}t}\pm{e^{-\lambda_{\mp}\tau}\over\sqrt\kappa}\langle
f_R(t)\alpha_{\pm}(0)\rangle$$
$$+{e^{-\lambda_{\mp}\tau}\over\sqrt\kappa}\int_0^t e^{-\lambda_{\mp}(t-t')}\bigg[\langle
f_R^*(t)f^*(t')\rangle\pm\langle f_R^*(t)f(t')\rangle\bigg]dt'$$
$$+{e^{-\lambda_{\mp}\tau}\over\sqrt\kappa}\int_0^\tau e^{\lambda_{\mp}\tau'}\bigg[\langle
f_R^*(t)f^*(t+\tau')\rangle\pm\langle f_R(t)f^*(t+\tau')\rangle$$
$$\pm\langle f_R^*(t)f(t+\tau')\rangle+\langle
f_R(t)f(t+\tau')\rangle\bigg]d\tau'.\eqno(3.49)$$Assuming the
noise force at time t does not affect the cavity mode variable at
an earlier times and taking into account the fact that
$$\langle f_R(t)\rangle=0,\eqno(3.50)$$we see that
$$\langle \alpha _{\pm}(0)f_R(t)\rangle=\langle \alpha _{\pm}(0)\rangle\langle
f_R(t)\rangle=0.\eqno(3.51)$$On account of this result, Eq. (3.49)
reduces to
$$\langle\alpha_{in\pm}(t)\alpha_{\pm}(t+\tau)\rangle={e^{-\lambda_{\mp}\tau}\over\sqrt\kappa}\int_0^t e^{-\lambda_{\mp}(t-t')}\bigg[\langle
f_R^*(t)f^*(t')\rangle\pm\langle f_R^*(t)f(t')\rangle\bigg]dt'$$
$$+{e^{-\lambda_{\mp}\tau}\over\sqrt\kappa}\int_0^\tau e^{\lambda_{\mp}\tau'}\bigg[\langle
f_R^*(t)f^*(t+\tau')\rangle\pm\langle f_R(t)f^*(t+\tau')\rangle$$
$$\pm\langle f_R^*(t)f(t+\tau')\rangle+\langle
f_R(t)f(t+\tau')\rangle\bigg]d\tau'.\eqno(3.52)$$The noise force
is a sum of the system and reservoir noise forces:
$$f(t)=f_R(t)+f_S(t),\eqno(3.53)$$where ~$f_R(t)$~and~$f_S(t)$~are the
reservoir and system noise forces. With the aid of this, (3.52)
can be rewritten as
$$\langle\alpha_{in\pm}(t)\alpha_{\pm}(t+\tau)\rangle={e^{-\lambda_{\mp}\tau}\over\sqrt\kappa}\int_0^t e^{-\lambda_{\mp}(t-t')}[\langle
f_R^*(t)f_R^*(t')\rangle+\langle f_R^*(t)f_S^*(t')\rangle$$
$$\pm\langle f_R^*(t)f_R(t')\rangle\pm\langle f_R^*(t)f_S(t')\rangle]dt'$$
$$+{e^{-\lambda_{\mp}\tau}\over\sqrt\kappa}\int_0^\tau e^{\lambda_{\mp}\tau'}[\langle
f_R^*(t)f_R^*(t+\tau')\rangle+\langle
f_R^*(t)f_S^*(t+\tau')\rangle$$
$$\pm\langle f_R(t)f_R^*(t+\tau')\rangle\pm\langle f_R(t)f_S^*(t+\tau')\rangle$$
$$\pm\langle f_R^*(t)f_R(t+\tau')\rangle\pm\langle f_R^*(t)f_S(t+\tau')\rangle$$
$$+\langle f_R(t)f_R(t+\tau')\rangle+\langle f_R(t)f_S(t+\tau')\rangle]d\tau'.\eqno(3.54)$$
With the aid of (3.53), Eq. (2.32) can be rewritten as
$$\langle f_R(t)f_R(t')\rangle+\langle f_R(t)f_S(t')\rangle+\langle
f_S(t)f_R(t')\rangle+\langle
f_S(t)f_S(t')\rangle=-2v\delta(t-t'),\eqno(3.55a)$$in which
$$v={A\over 2B}\bigg[-\rho_{aa}^{(0)}\left({\Omega\over
\gamma}\right)\left(1+{\Omega^2\over
4\gamma^2}\right)+\rho_{cc}^{(0)}\left({\Omega\over
2\gamma}\right)\left(1-{\Omega^2\over 2\gamma^2}\right)$$
$$-\rho_{ac}^{(0)}\left(1-{\Omega^2\over
2\gamma^2}\right)\bigg]-{\kappa \over 2}M,\eqno(3.550b).$$ Since
the reservoir and the system noise forces are uncorrelated, we see
that
$$\langle f_R(t)f_S(t')\rangle=\langle f_R(t)\rangle\langle
f_S(t')\rangle=0,\eqno(3.56a)$$
$$\langle f_S(t)f_R(t')\rangle=\langle f_S(t)\rangle\langle
f_R(t')\rangle=0.\eqno(3.56b)$$Thus Eq. (3.55a) reduces to
$$\langle f_R(t)f_R(t')\rangle+\langle
f_S(t)f_S(t')\rangle=-2v\delta(t-t'),\eqno(3.57a)$$from which
follows
$$\langle f_R(t)f_R(t')\rangle=\kappa M\delta(t-t'),\eqno(3.57b)$$
$$\langle f_S(t)f_S(t')\rangle={A\over 2B}\bigg[-\rho_{aa}^{(0)}\left({\Omega\over
\gamma}\right)\left(1+{\Omega^2\over
4\gamma^2}\right)+\rho_{cc}^{(0)}\left({\Omega\over
2\gamma}\right)\left(1-{\Omega^2\over 2\gamma^2}\right)$$
$$-\rho_{ac}^{(0)}\left(1-{\Omega^2\over
2\gamma^2}\right)\bigg]\delta(t-t').\eqno(3.57c)$$

Moreover employing (3.53), Eq. (2.35) can be rewritten as
$$\langle f_R(t')f_R^*(t)\rangle+\langle f_R(t')f_S^*(t)\rangle+\langle
f_S(t')f_R^*(t)\rangle+\langle
f_S(t')f_S^*(t)\rangle=2p\delta(t-t'),\eqno(3.58a)$$in which
$$p={A\over 2B}\left[\rho_{aa}^{(0)}\left(1+{\Omega^2\over
4\gamma^2}\right)+\rho_{cc}^{(0)}\left({3\Omega^2\over
4\gamma^2}\right)-\rho_{ac}^{(0)}\left({3\Omega\over
2\gamma}\right)\right]+{\kappa \over 2}N.\eqno(3.58b)$$ Using the
fact that
$$\langle f_R(t')f_S^*(t)\rangle=\langle
f_S(t')f_R^*(t)\rangle=0,\eqno(3.59a)$$we observe that
$$\langle f_R(t')f_R^*(t)\rangle=\kappa
N\delta(t-t'),\eqno(3.59b)$$
$$\langle f_S(t')f_S^*(t)\rangle={A\over 2B}\left[\rho_{aa}^{(0)}\left(1+{\Omega^2\over
4\gamma^2}\right)+\rho_{cc}^{(0)}\left({3\Omega^2\over
4\gamma^2}\right)-\rho_{ac}^{(0)}\left({3\Omega\over
2\gamma}\right)\right]\delta(t-t').\eqno(3.59c)$$
 On account of  Eqs. (3.56), (3.57b), (3.59a), and (3.59b), Eq. (3.54) takes the form
$$\langle\alpha_{in\pm}(t)\alpha_{\pm}(t+\tau)\rangle=2\sqrt\kappa(M\pm
N)e^{-\lambda_{\mp}\tau}.\eqno(3.60)$$

Furthermore, on account of (3.39b), one can write
$$\langle\alpha_{in\pm}(t)\alpha_{in\pm}(t+\tau)\rangle={1\over\kappa}[\langle
f_R^*(t)f_R^*(t+\tau)\rangle\pm\langle
f_R^*(t)f_R(t+\tau)\rangle\pm\langle
f_R(t)f_R^*(t+\tau)\rangle+\langle
f_R(t)f_R(t+\tau)\rangle],\eqno(3.61)$$so that using Eqs. (3.57b)
and (3.59b), one gets
$$\langle\alpha_{in\pm}(t)\alpha_{in\pm}(t+\tau)\rangle=2(M\pm
N)\delta(\tau).\eqno(3.62)$$ Now with the aid of (3.44b),
(3.47), (3.60), and (3.62), the squeezing spectrum can be put in
the form
$$S_{\pm}^{out}(\omega)=1\pm 2\kappa
Re\int_0^{\infty}\left({-2v\pm
2p\over\lambda_{\mp}}\right)~e^{-\lambda_{\mp}\tau+i\omega\tau}d\tau$$
$$\mp 2\sqrt\kappa Re\int_0^{\infty}2\sqrt\kappa(M\pm N)~e^{-\lambda_{\mp}\tau+i\omega\tau}d\tau$$
$$\pm 2 Re\int_0^{\infty}2(M\pm
N)\delta(\tau)~e^{i\omega\tau}d\tau.\eqno(3.63)$$Upon performing
the integration and taking into account Eqs. (3.14a), (3.14c) and
(3.17b), we get
$$S_{\pm}^{out}(\omega)=1+{4\kappa(p\mp
v)\mp{4\kappa(M\pm
N)\lambda_{\mp}\over{\omega^2+\lambda_{\mp}^2}}}+2N\pm
2M,\eqno(3.64)$$where
$$(p\mp v)={\kappa\over 2}(M\pm N)+{A\over
4B}\bigg[(1+\beta^2)(1\pm{\beta\over 2})+\eta({\beta\over
2}(\beta\mp3)-1\bigg]$$
$$+{A\sqrt{1-\eta^2}\over 4B}\bigg[-{\beta\over 2}(3\pm\beta)\pm1\bigg],\eqno(3.65a)$$
$$B=(1+\beta^2)(1+{\beta^2\over 4}),\eqno(3.65b)$$and
$$\lambda_{\mp}={\kappa\over 2}+{A\over
4B}\bigg[\eta(2-\beta^2)\mp\beta(1+\beta^2)+\sqrt{1-\eta^2}(3\beta)\bigg].\eqno(3.65c)$$Setting~$\eta=0$~and
taking into account Eqs. (2.8f) and (2.8g), we have
$$S^{out}_+=e^{2r}+{4\kappa(p-v)-2\kappa(e^{2r}-1)\lambda_{-}\over{\omega^2+\lambda_{-}^2}}
,\eqno(3.66a)$$and
$$S^{out}_-=e^{-2r}+{4\kappa(p+v)+2\kappa(1-e^{-2r})\lambda_{+}\over{\omega^2+\lambda_{+}^2}},\eqno(3.66b)$$where
$$p-v={\kappa\over 4}(e^{2r}-1)+{A(\beta^3+\beta^2-2\beta+4)\over(1+\beta^2)(2+{\beta^2\over
2})},\eqno(3.67a)$$
$$p+v={\kappa\over 4}(e^{-2r}-1)-{A(\beta^3-3\beta^2+4\beta)\over(1+\beta^2)(2+{\beta^2\over
2})},\eqno(3.67b)$$
$$\lambda_-={\kappa\over 2}\bigg(1+{A\beta(2-\beta^2)\over 2\kappa(1+\beta^2)(1+{\beta^2\over
4})}\bigg),\eqno(3.67c)$$
$$\lambda_+={\kappa\over 2}\bigg(1+{2A\beta\over
\kappa(1+\beta^2)}\bigg).\eqno(3.67d)$$
\begin{center}
\begin{figure}[h]
\centerline{\includegraphics [width=2.5 in]{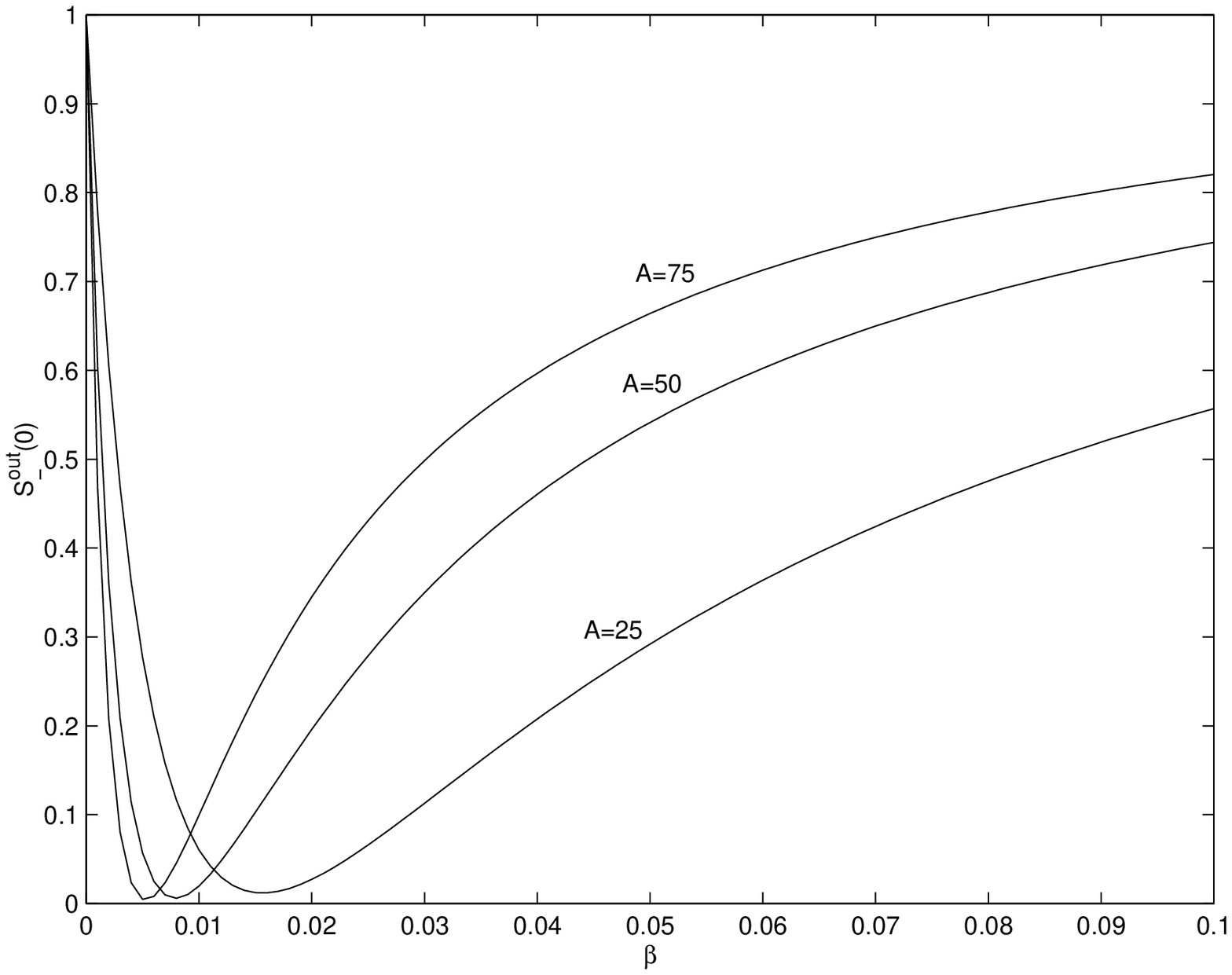}} {Fig.
3.7 {\footnotesize Plots of squeezing spectrum
versus~$\beta$~for~$r=0,$~$\kappa=0.8,$~$\eta =0,$~ ~$\omega
=0$~and for different values of the linear gain coefficient .}}
\end{figure}
\end{center}

\begin{center}
\begin{figure}[h]
\centerline{\includegraphics [width=2.5 in]{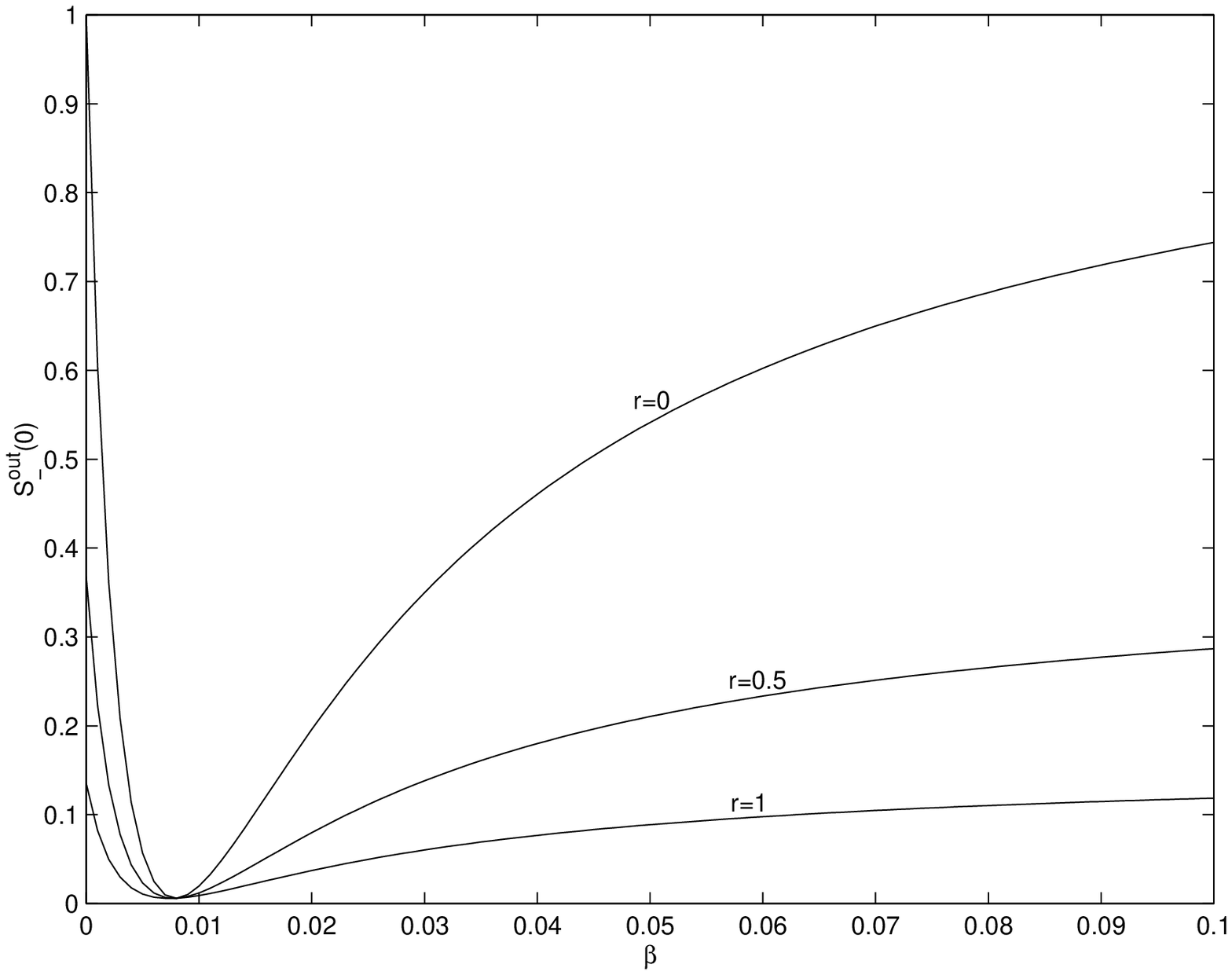}} {Fig.
3.8 {\footnotesize Plots of squeezing spectrum
versus~$\beta$~for~$A=50,$~$\kappa=0.8,$~$\omega =0,$~~$\eta
=0$~and for different values of the squeeze parameter .}}
\end{figure}
\end{center}
of the output mode increases with  linear gain coefficient~$A$~ or
the squeeze parameter~$r.$ In addition, both figures show that
there is almost perfect squeezing for small value of~$\beta.$~

\section{Photon Statistics}

 In this section we wish to calculate the
mean and the variance of the photon number for the cavity mode
under consideration.
\subsection{ The Mean Photon Number}
The mean photon number can be expressed in terms of c-number
variable associated with the normal ordering as
$$\langle
\hat{n}\rangle=\langle\alpha^*(t)\alpha(t)\rangle.\eqno(4.1)$$Making
use of Eq. (2.39) along with its complex conjugate, we obtain
$$\langle\alpha^*(t)\alpha(t)\rangle=A^2(t)\langle\alpha^*(0)\alpha(0)\rangle+B^2(t)\langle\alpha(0)\alpha(0)^*\rangle$$
$$+A(t)B(t)[\langle\alpha^2(0)\rangle+\langle\alpha^{*2}(0)\rangle]$$
$$+A(t)[\langle\alpha^*(0)F(t)\rangle+\langle
F^*(t)\alpha(0)\rangle$$
$$+B(t)[\langle\alpha(0)F(t)\rangle+\langle
F^*(t)\alpha^*(0)\rangle$$
$$+\langle F^*(t)F(t)\rangle,\eqno(4.2)$$where A(t), B(t), and
F(t) are given by Eq. (2.40). Employing Eq. (2.30) together with
the fact that the cavity mode is initially in a vacuum state, one
obtains
$$\langle\alpha^*(t)\alpha(t)\rangle=\langle
F^*(t)F(t)\rangle.\eqno(4.3)$$ On account of  Eqs. (2.40c)
and (2.40d), we can write
$$\langle F^*(t)F(t)\rangle={1\over 4}e^{-2\lambda_-t}\int_0^t
e^{\lambda_-(t'+t'')}[\langle f^*(t'')f(t')\rangle+\langle
f^*(t'')f^*(t')\rangle$$
$$+\langle f(t'')f(t')\rangle+\langle
f(t'')f^*(t')\rangle]dt'dt''$$
$$+{1\over 4}e^{-(\lambda_-+\lambda_+)t}\int_0^te^{\lambda_-t''+\lambda_+t'}[\langle f^*(t'')f(t')\rangle-\langle
f^*(t'')f^*(t')\rangle$$
$$+\langle f(t'')f(t')\rangle-\langle
f(t'')f^*(t')\rangle]dt'dt''$$
$$+{1\over 4}e^{-(\lambda_-+\lambda_+)t}\int_0^te^{\lambda_-t''+\lambda_+t'}[\langle f^*(t'')f(t')\rangle+\langle
f^*(t'')f^*(t')\rangle$$
$$-\langle f(t'')f(t')\rangle-\langle
f(t'')f^*(t')\rangle]dt'dt''$$
$$+{1\over 4}e^{-2\lambda_+t}\int_0^t
e^{\lambda_+(t'+t'')}[\langle f^*(t'')f(t')\rangle-\langle
f^*(t'')f^*(t')\rangle$$
$$-\langle f(t'')f(t')\rangle+\langle
f(t'')f^*(t')\rangle]dt'dt''.\eqno(4.4)$$Applying Eqs. (2.32) and
(2.35), Eq. (4.4) can be put in the form
$$\langle
F^*(t)F(t)\rangle=(p-v)e^{-2\lambda_-t}\int_0^te^{\lambda_-(t'+t'')}\delta(t''-t')dt'dt''$$
$$+(p-v)e^{-2\lambda_+t}\int_0^te^{\lambda_+(t'+t'')}\delta(t''-t')dt'dt''.\eqno(4.5)$$On
performing the integration using the property of Dirac delta
function
$$\int_0^tf(t'')~\delta(t''-t')dt''=f(t'),\eqno(4.6)$$one readily
finds
$$\langle
F^*(t)F(t)\rangle=\left[{p-v\over2\lambda_-}\right](1-e^{-2\lambda_-t})
+\left[{p+v\over2\lambda_+}\right](1-e^{-2\lambda_+t}).\eqno(4.7)$$With
the aid of Eqs. (4.3), (2.37b) ,(2.17a) and (2.17b), the mean
photon number of the cavity mode at steady state turns out to be
$$\bar{n}_{ss}={p(q-p)-v(u-v)\over(p-q)^2-(u-v)^2}.\eqno(4.8)$$
\subsection{The Variance of the Photon Number}
The variance of the photon number for the cavity mode is given in
the normal order  by
$$\Delta{n}^2(t)=\langle\hat{a}^{\dag2}(t)\hat{a}^2(t)\rangle+\langle\hat{a}^\dag(t)\hat{a}(t)\rangle-\langle\hat{a}^\dag(t)\hat{a}(t)\rangle^2.\eqno(4.9)$$The
c-number equation corresponding to (4.9) is
$$\Delta{n}^2(t)=\langle\alpha^*(t)\alpha(t)\rangle+\langle\alpha^{*2}(t)\alpha^2(t)\rangle-\langle\alpha^*(t)\alpha(t)\rangle^2.\eqno(4.10)$$Since
the stochastic differential equation for ~$\alpha(t)$~has the form
given by Eq. (2.18) and noise operator f(t) has the correlation
properties described by Eqs. (2.20), (2.32) and (2.35),
~$\alpha$~is a Gaussian variable. Therefore one can write [2]
$$\langle\alpha^{*2}(t)\alpha^2(t)\rangle=\langle\alpha^{*2}(t)\rangle\langle\alpha^2(t)\rangle+2\langle\alpha^{*}(t)\alpha(t)\rangle^2,\eqno(4.11)$$so
that Eq. (4.10) can be expressed as
$$\Delta{n}^2(t)=\langle\alpha^{*2}(t)\rangle\langle\alpha^2(t)\rangle+\langle\alpha^{*}(t)\alpha(t)\rangle^2+\langle\alpha^{*}(t)\alpha(t)\rangle.\eqno(4.12)$$With
the aid of  Eqs. (2.39) and (2.40), one readily obtains
$$\langle\alpha^2(t)\rangle=\left[{p-v\over2\lambda_-}\right](1-e^{-2\lambda_-t})
-\left[{p+v\over2\lambda_+}\right](1-e^{-2\lambda_+t}).\eqno(4.13)$$And
at steady state
$$\langle\alpha^2(t)\rangle_{ss}=\langle\alpha^{*2}(t)\rangle_{ss}={p(u-v)-v(q-p)\over(q-p)^2-(u-v)^2}.\eqno(4.14)$$Hence
on account of Eqs. (4.12) together with (4.8) and (4.14), the
variance of the photon number for the cavity mode, at steady
state, can be expressed as
$$\Delta{n}^2_{ss}=\bar{n}_{ss}^2+\bar{n}_{ss}+\left[\bar{n}_{ss}-{(p+v)\over(q+u)-(p+v)}\right].\eqno(4.15)$$

\begin{center}
\begin{figure}[h]
\centerline{\includegraphics [width=2.5 in]{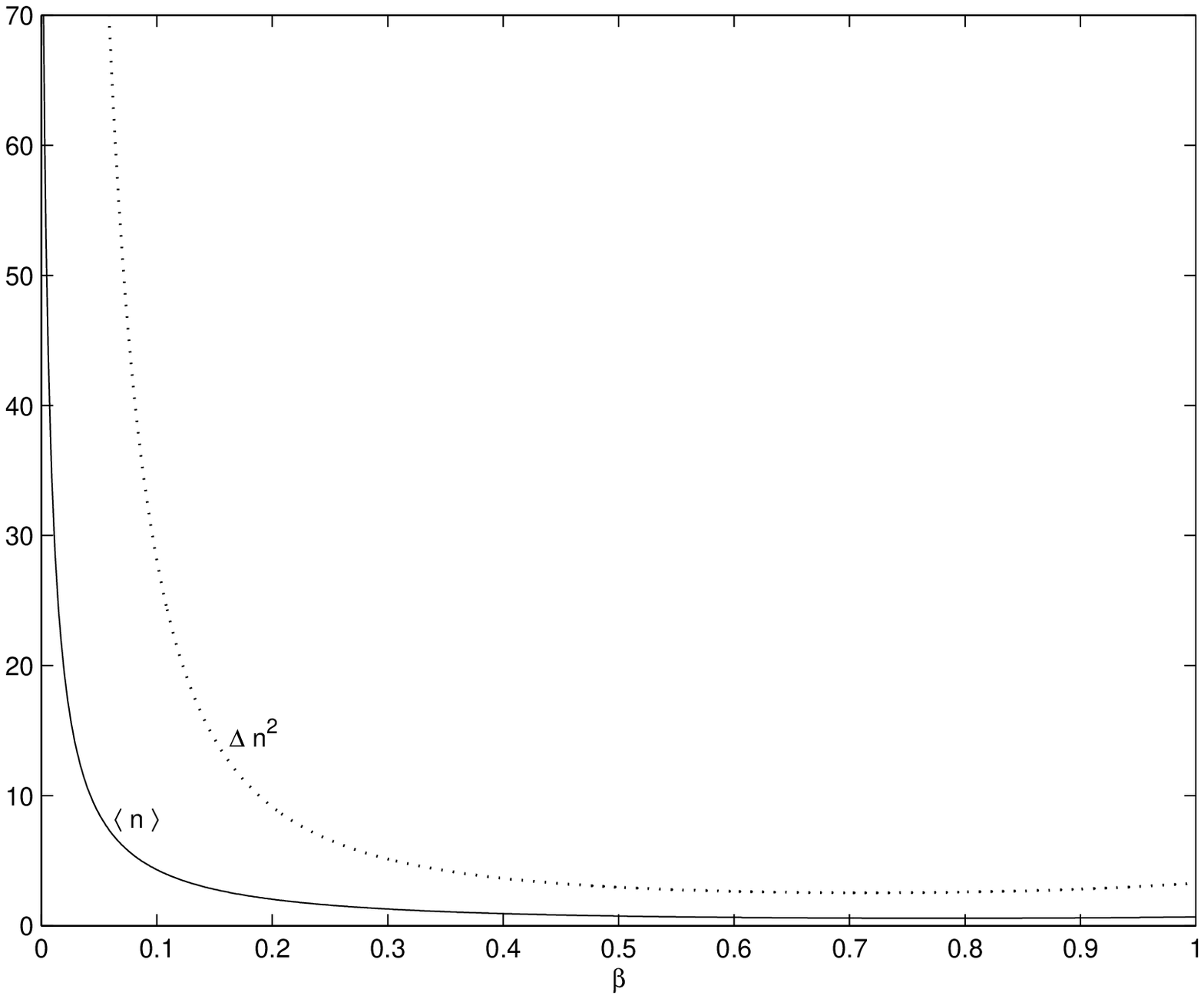}} {Fig.
3.9 {\footnotesize   Plots of mean photon number and variance of
the photon number
versus~$\beta$~for~$A=75,$~$\kappa=0.8,$~\\$r=0.$~}}
\end{figure}
\end{center}


Fig. (3.9) shows that the mean photon number decreases
as~$\beta$~increases. We also note that the photon statistics of
the cavity mode under consideration is super-Poisssonian.
\section{Conclusion}
\label{sec:conclusion}
We have considered a three-level laser
coupled to a squeezed vacuum reservoir and in which the top and
bottom levels are coupled by a strong coherent light. Employing
the master equation, for the cavity mode under consideration, we
have obtained stochastic differential equations. Applying the
solutions of these equations, we have calculated the quadrature
variance and squeezing spectrum. From the plots of the quadrature
variance versus~$\eta,$~we have seen that the cavity mode under
consideration is in a squeezed state for all values
of~$\eta$~between zero and one and the degree of squeezing
increases with the linear gain coefficient or the the squeeze
parameter. We have also seen from the plots of the quadrature
variance versus~$\beta,$~ when all the atoms are initially in the
upper level or when half of the atoms are initially in the upper
level with the remaining half being in the lower level and when
these levels are coupled by a strong coherent light, that the
cavity mode is in a squeezed state for small values of~$\beta$~and
the degree of squeezing increases with the linear gain coefficient
or the squeeze parameter.

Moreover, the plots of the squeezing spectrum
versus~$\beta$~for~$\eta=0$~ show that there is almost perfect
squeezing for large values of the linear gain coefficient or the
squeeze parameter.
 In addition, applying the
solutions of the stochastic differential equations, we have
calculated the mean photon number and the variance of the photon
number for the cavity mode. We see that the photon statistics of
the cavity mode is super-Poisssonian. Fig. (3.9) shows that the
mean photon number decreases as~$\beta$~increases.

\section{References}

\end{document}